\documentclass[universe,article,accept,pdftex,moreauthors]{Definitions/mdpi} 

\newcommand{\Mp}{M_{\rm Pl}}

\newcommand{\xc}{x_{\rm c}}

\newcommand{\Qtwo}{\ensuremath{\mathcal{Q}_{2}}}
\newcommand{\phitwo}{\ensuremath{\phi_{2}}}
\newcommand{\rhos}{\ensuremath{\rho_{\rm s}}}
\newcommand{\rs}{\ensuremath{r_{\rm s}}}
\newcommand{\Cs}{\ensuremath{C_{\rm s}}}

\newcommand{\dr}[1] {\frac{\text{d}#1}{\text{d}r}}

\newcommand{\apj}{Astrophys. J.}

\newcommand{\mnras}{Mon. Not. R. Astron. Soc.}
\newcommand{\aap}{Astron. Astrophys.}
\newcommand{\prd}{Phys.\ Rev.\ D}

\firstpage{1} 
\pubvolume{1}
\issuenum{1}
\articlenumber{0}
\pubyear{2026}
\copyrightyear{2026}
\externaleditor{Firstname Lastname} 
\datereceived{7 March 2026} 
\daterevised{13 April 2026} 
\dateaccepted{24 April 2026} 
\datepublished{ } 

\Title{CLASH-VLT: The Fifth Force in Chameleon Gravity from Joint Lensing and Kinematics Cluster Mass Profiles}

\Author{Lorenzo 
 Pizzuti $^{1,2,}${*}
\orcidA{}, Federico Rivano $^{1}$, Keiichi Umetsu $^{3}$ and Andrea Biviano $^{2,4}$}

\AuthorNames{Lorenzo Pizzuti, Federico Rivano, Keiichi Umetsu and Andrea Biviano}

\address{%
$^{1}$ \quad Dipartimento di Fisica G. Occhialini, Universit\`a di Milano-Bicocca,
Piazza della Scienza 3, 20126 Milano, Italy; {f.rivano@campus.unimib.it} 
\\
$^{2}$ \quad INAF-Osservatorio Astronomico di Trieste, via G. Tiepolo 11, {34143} 
 Trieste, Italy; {andrea.biviano@inaf.it}\\

$^{3}$ \quad Academia Sinica Institute of Astronomy and Astrophysics (ASIAA), No. 1, Section~4, Roosevelt Road, \linebreak{}Taipei 106216, Taiwan; {keiichi@asiaa.sinica.edu.tw}\\

$^4$ \quad IFPU---Institute for Fundamental Physics of the Universe, via Beirut 2, 34014 Trieste, Italy\\

}

\corres{Correspondence: lorenzo.pizzuti@unimib.it}

\abstract{
We present a high-precision joint gravitational-lensing and kinematic analysis of nine massive galaxy clusters from the CLASH and CLASH-VLT surveys to test chameleon screening gravity and its $f(R)$ sub-class at Mpc scales. We investigate the dependence on the assumed parametrization of the total cluster mass profile by adopting three models, namely Navarro--Frenk--White (NFW), Burkert, and Hernquist. When cuspy models (NFW or Hernquist) are assumed in the general chameleon framework, the combined constraints from the nine clusters are fully consistent with General Relativity (GR), excluding large regions of the modified-gravity parameter space (the coupling constant $\mathcal{Q}$ and the background chameleon field $ \phi_\infty$), providing one of the tightest bounds on general chameleon models with clusters to date. 
In contrast, 
 adopting a Burkert profile---disfavored by lensing data---leads to a mild ($\sim 2\sigma$) departure from the GR expectation in joint analysis. 
When considering the $f(R)$ sub-case,  we obtain a bound on the background scalaron field of $|f_R| \lesssim \mathrm{2-5}\times 10^{-5}$ (95\% C.L.) for NFW and Hernquist models, in agreement with current constraints at cosmological scales, and an apparent deviation from standard gravity of $\log_{10}|f_R| = -4.7 \pm 1.2$ for the Burkert case. We investigate the impact of systematics in the kinematical analysis, showing that the tension is mitigated when clusters exhibiting clear dynamical disturbance are excluded from the sample.
Our results show that galaxy clusters provide competitive tests of screened modified gravity at mega-parsec scales, while highlighting the critical role of accurate mass modeling and dynamical-state assessment. The upcoming generation of wide-field lensing surveys and spectroscopic follow-up programs will enable similar analyses on substantially larger samples, offering the prospect of tightening cluster-based constraints on gravity and the dark sector.}


\begin{document}


\section{Introduction}

Since its discovery, the~origin of the low-redshift accelerated expansion of the Universe~\cite{2016ApJ...833L..30R,2012CRPhy..13..521A} has became one of the fundamental open questions in cosmology. In~the current cosmological paradigm, formalized by the so-called $\Lambda$CDM model, this~acceleration arises from   
a \empty{{cosmological constant}
}, $\Lambda$, in~Einstein's field equations of General Relativity (GR, hereafter), for which standard physics has found no natural explanation 
 ({{e.g.,}
}~\cite{Martin2012,Padilla15}). In~the simplest interpretation, the~cosmological constant encapsulates the contribution of a new unknown component accounting for $\sim$68\% of the energy budget of the universe~\cite{Planck2020}, dark energy (DE) 
 \cite{DE2024}, characterized by a constant equation of state $p_\Lambda = w\rho_\Lambda$, with~$w = -1$.

While the $\Lambda$CDM model remarkably describes a large variety of phenomena across different scales, the~increase in the quality and quantity of datasets has highlighted several tensions between early and late time observations (see, {e.g.,}~\cite{Perivolaropoulos_2022,Abbott_2022}). 
{In addition, recent {{DESI}} results~\cite{2024arXiv240413833W} have provided evidence in favor of a dynamical DE evolving with cosmic time $w = w(z)$, lowering the efficacy of the $\Lambda$CDM model even more and suggesting the need for some modification. However, more recent Bayesian studies~\cite{ong2026} have shown that DESI DR2's 
	 deviation from $\Lambda$CDM is largely driven by inter-dataset tensions, and it may be reduced once residual systematic effects are properly accounted for. This highlights the subtle 
	  interplay among systematics, dataset consistency, and theoretical modeling when testing non-standard cosmological scenarios.}

Among the various possibilities, a~large-scale remodeling of the gravitational interaction (i.e.,~modifying GR) is also being taken in consideration. While GR has been tested with enormous success on small scales~\cite{Everitt1980,2017arXiv170504397A,Ciufolini2024}, modifications at a cosmological level may still be viable and provide a phenomenology which mimics the background evolution of the $\Lambda$CDM paradigm~\cite{Koyama:2015oma,CANTATA:2021asi}. Both dark sector and modified gravity (MG) alternatives change the formation and evolution of cosmic structures through the introduction of new (dynamical) degrees of freedom. 
In this regard, galaxy clusters emerge as exceptional laboratories to constrain the $\Lambda$CDM {framework} \cite{Haggar24} and possible signatures of new physics at a cosmological level~\cite{Sakstein:2015aac,Pizzuti17,Artis25}. In~fact, they are the largest and most massive self-gravitating systems in the universe, constituting the endpoint of the structure formation process~\cite{Kravtsov12}. 

The internal shape of a cluster and the distribution of both baryonic and dark matter carry pivotal information on the nature of gravity and of the dark sector~\cite{2015MNRAS.449.2837G,2024A&A...690A.364B}. Indeed, MG or new dynamical components change the relation between matter perturbation and gravitational potentials. Non-relativistic (galaxies and gas) and relativistic (photons) tracers of the mass profile respond differently to gravitational interaction if {GR} is modified; thus, joint analyses combining lensing~\cite{2013SSRv..177...75H} and kinematics~\cite{Biviano2013} mass reconstruction can be used to constrain MG models at Mpc scales.
Since GR predictions fit Solar System and galactic tests almost perfectly, in~order to preserve their efficacy, viable MG theories need to implement a screening mechanism~\cite{2013CQGra..30u4005B,2016RPPh...79d6902K,2021Univ....8...11B}, which suppresses the new degrees of freedom and restores standard gravity at small scales. The~details of the screening—namely how efficiently the suppression is achieved 
—determine the resulting phenomenology, imprinting distinctive signatures on the mass distribution and dynamical properties of galaxy clusters ({e.g.,}~\cite{PizzutiSOLO}).

In this work, we focus on the popular {{chameleon screening}} theories~\cite{Khoury_2004,Brax_2010,Burrage:2017shh}---a branch of scalar--tensor theories of gravity---where the new degree of freedom generates a fifth force affecting the motion of non-relativistic particles, which depends on the matter density itself.  If~appropriately tuned, in~high-density regions the resulting effective mass of the chameleon field becomes so large that the fifth-force results are negligible 
 and GR is {re-established}. The~strength of the modification depends on the background value of the chameleon field $\phi_\infty$ and the coupling with matter $\mathcal{Q}$.
A widely studied sub-class of chameleon models is the $f(R)$ gravity~\cite{2007PhRvD..75j4016B,2007PhRvD..76f3504F}. In~this framework the Einstein--Hilbert action is generalized, introducing a function of the Ricci scalar, $f(R)$. As~a result, the~geometry of spacetime is modified, and~an additional fifth force is mediated by the scalaron field, $f_R = \partial f/\partial R$, which can be mapped onto the chameleon field.
A large number of stringent constraints has been obtained for both general chameleon models and specific $f(R)$ realizations across a wide range of physical scales~\cite{2016JCAP...11..045B,2018LRR....21....1B,Desmond2020,Benisty2023,Wilcox:2015kna,Cataneo:2016iav,Tamosiunas_2022}. 

{Recent works have further explored general screened modified gravity using strong-lensing observables on galactic and sub-galactic scales. In~particular,~\cite{Jyoti2019} introduced a phenomenological parametrization of MG in terms of a post-Newtonian slip parameter $\gamma_{\rm PN}$ and a screening scale $\Lambda$, modeling the transition from GR to a modified regime as a sharp change in the relation between the metric potentials. Using time delays in strong lenses, they obtained constraints of the form $|\gamma_{\rm PN}-1| \lesssim 0.2 \, (\Lambda / 100\,{\rm kpc})$, effectively probing scales in the range $\Lambda \sim 10$--$200\,{\rm kpc}$. Similar approaches have been adopted in galaxy-scale analyses combining strong lensing and stellar dynamics to constrain deviations from GR in a largely phenomenological framework (e.g.,~\cite{Lian2022}).}

{More recently, strong lensing of gravitational waves has been proposed as a novel probe of screened MG models by constraining $\gamma_{\rm PN}$ (e.g.,~\cite{Mu2026}). These analyses provide a new window on gravity at large scales, being sensitive to the relativistic sector and to the behavior of gravitational waves across cosmological distances.}

{These approaches are complementary to the present work, but~conceptually different. First, chameleon-like theories are characterized by a screening mechanism that depends primarily on the local gravitational potential (or density), rather than on a fixed physical scale. In~this case, the~transition between screened and unscreened regimes can be associated with an effective screening radius $S$, defined implicitly by the structure of the density field, typically of $\mathcal{O}({\rm Mpc})$, making possible deviations hard to detect in galactic environments. 
In this regard, galaxy clusters remain a particularly interesting regime to explore; they lay at the interface between astrophysics and cosmology, where screening mechanisms may operate in a non-trivial and potentially observable way, leaving characteristic imprints on cluster mass profiles. Second, while strong-lensing analyses probe the relativistic sector typically on galaxy scales, the~analysis presented in this work exploits a joint comparison between relativistic (strong {and} weak lensing) and non-relativistic (galaxy kinematics) tracers at Mpc scales. This allows us to directly constrain the physical parameters of chameleon gravity, namely the coupling $\mathcal{Q}$ and the background field value $\phi_\infty$, extending existing constraints to environments where the chameleon mechanism is expected to operate differently due to the diverse densities and potentials involved.}

Moreover, despite its relative simplicity, chameleon gravity provides a controlled yet versatile theoretical framework that makes it an interesting case study to fully exploit the constraining power of clusters, and~to develop methodologies that can later be extended to more complex MG~scenarios. 

In the following, we use high precision gravitational lensing and member galaxies' kinematical information of nine massive galaxy clusters, extensively studied within the CLASH~\cite{Postman12} and CLASH-VLT~\cite{2014Msngr.158...48R} collaborations, in~order to constrain the general chameleon gravity and the sub-class of $f(R)$ models. By~means of the \textsc{MG-MAMPOSSt} code~\cite{Pizzuti:2020tdl}, we obtain the joint lensing-kinematics posterior distributions of the free parameters describing the chameleon model to constrain the allowed region of the parameter space. 
Since the efficiency of the chameleon screening is strongly related to the structure of the matter distribution, we explore how different {prescriptions} of the total cluster mass profile affect the results. {This follows from the theoretical analysis performed in~\cite{Pizzuti2024b}, where we implemented semi-analytical solutions of chameleon gravity for different mass ansatze and we validated them against simulated halos; here, we apply such a framework to real cluster datasets for the first time.} We further discuss the impact of systematics, in~particular deviation dynamical relaxation and the presence of substructures in the distribution of the member~galaxies. 

This paper is structured as follows: in Section~\ref{sec:theo} we provide a brief overview of the chameleon model; in~Section~\ref{sec:data} we present the dataset and the methodology used for the analysis before presenting the main results in Section~\ref{sec:results}. Finally, we highlight our main conclusions in Section~\ref{sec:conclusions}.

\section{Theoretical~Background}
\label{sec:theo}


Scalar--tensor theories with screening mechanisms provide a well-motivated
framework to reconcile modifications of gravity on cosmological scales with
stringent local constraints. Among~them, the~chameleon {formalism}
\cite{Khoury_2004} represents a prototypical example, in~which the scalar
degree of freedom dynamically adapts its mass to the surrounding environment,
suppressing deviations from GR in high-density regions
while remaining active on large~scales.

The dynamics of a chameleon scalar field $\phi$, conformally coupled to matter
fields $\psi^{(i)}$, is described by the Einstein-frame Lagrangian
\begin{equation}
\label{eq:lagrangian}
\mathcal{L} = \sqrt{-g}\left[
-\frac{M_{\rm Pl}^{2}}{2}R
+ \frac{1}{2}(\partial\phi)^2
+ V(\phi)
\right]
+ \mathcal{L}_{\rm m}\!\left(\psi^{(i)}, g^{(i)}_{\mu\nu}\right),
\end{equation}
where $\Mp=(8\pi G)^{-1/2}$ is the reduced Planck mass, and~$G$ is the gravitational constant.
Matter fields follow geodesics of the Jordan-frame metric
\begin{equation}
g^{(i)}_{\mu\nu}
= e^{-2 Q_i \phi/(M_{\rm Pl} c^2)}\, \tilde g_{\mu\nu},
\end{equation}
with $Q_i$ denoting the dimensionless coupling between the scalar field and the
$i$-th matter species, and~$c$ being the speed of light. Throughout this work, we adopt a universal coupling
$Q$ for both baryonic and dark matter components, as~commonly assumed in
phenomenological analyses
({e.g.,}~\cite{Wilcox:2015kna,Butt:2024jes}).

A particularly relevant case is provided by metric $f(R)$ theories of gravity,
which can be recast as scalar--tensor models with a chameleon-like screening
mechanism. In~this correspondence, the~scalar degree of freedom is identified
with $f_R \equiv \partial f/\partial R$, and~the coupling takes the fixed value
$Q=1/\sqrt{6}$ \cite{starobinsky2007disappearing,oyaizu2008nonlinear}. The~connection between the chameleon field and the scalaron is given by the following equation (see, for example,~\cite{Brax:2008}):
\begin{equation}\label{eq:conversion}
        |f_R| = \exp\left(-\frac{2\mathcal{Q}\phi}{\Mp c^2}\right)-1\ ,
    \end{equation}
where the quantity $\phi/\Mp$ carries the dimension of a gravitational potential.
As a
result, the~formalism we develop below applies equally to generic chameleon
models and to the class of viable $f(R)$ theories considered in this~work.

The self-interaction of the scalar field is encoded in a monotonic potential,
which can be described with an inverse power-law form (e.g.,~\cite{Burrage_2015}),
\begin{equation}
\label{eq:potential}
V(\phi) = \lambda^{4+n}\,\phi^{-n},
\end{equation}
where $n>0$ and $\lambda$ set the characteristic energy scale, typically
associated with the dark-energy scale~\cite{Tamosiunas_2022, Pizzuti2024b}.

In the quasi-static, non-relativistic regime relevant for galaxy clusters, the~scalar field equation of motion reads
\begin{equation}
\label{eq:campophi}
\nabla^2 \phi
= V'(\phi)
+ \frac{Q}{M_{\rm Pl}c^2}
\sum_j \rho_j\, e^{Q\phi/(M_{\rm Pl}c^2)},
\end{equation}
where the sum runs over all matter components with density $\rho_j$.
The coupling to matter gives rise to an effective potential,
\begin{equation}
\label{eq:effective_potential}
V_{\rm eff}(\phi)
= V(\phi)
+ \sum_j \rho_j\, e^{Q\phi/(M_{\rm Pl}c^2)},
\end{equation}
whose minimum depends explicitly on the ambient density.
Since current observational constraints require
$\phi/(M_{\rm Pl}c^2) \ll 1$
\cite{Xang19,Desmond2020,Boumechta:2023qhd},
the exponential term can be linearized, leading to
\begin{equation}
\label{eq:eq_of_motion_s}
\nabla^2 \phi
\simeq \frac{Q}{M_{\rm Pl}c^2}\sum_j \rho_j + V'(\phi).
\end{equation}
{In} 
 this form, $\phi$ plays the role of an additional gravitational potential,
mediating a fifth force, whose strength depends on the local density. In~particular, assuming spherical symmetry, the~total gravitational acceleration acting on non-relativistic tracers can be
written as
\begin{equation}
\label{eq:gradient}
\frac{\mathrm{d}\Phi}{\mathrm{d}r}
=
\frac{G M(r)}{r^2}
+
\frac{Q}{M_{\rm Pl}}
\frac{\mathrm{d}\phi}{\mathrm{d}r},
\end{equation}
which motivates the definition of an effective mass associated with the scalar
field,
\begin{equation}
M_{\rm eff}(r)
=
\frac{Q}{G}
\frac{r^2}{M_{\rm Pl}}
\frac{\mathrm{d}\phi}{\mathrm{d}r}.
\end{equation}
The total dynamical mass inferred from kinematic probes is therefore
\begin{equation}
\label{eq:massdyn}
M_{\rm dyn}(r)
=
M_{\rm GR}(r)
+
M_{\rm eff}(r)\,,
\end{equation}
where $M_{\rm GR}(r)$ is the total mass profile sourced by the matter components $\rho_j$.

An important feature of chameleon and $f(R)$ gravity is that null geodesics are
invariant under conformal transformations. As~a consequence, gravitational
lensing remains sensitive only to the Newtonian potential generated by
$M_{\rm GR}$ \cite{Burrage:2017shh}. Lensing observations, thus, provide an
independent and (ideally) unbiased probe of the true mass distribution, which we exploit
as a prior in our~analysis.
\subsection*{Semi-Analytical Solutions for the Chameleon~Field}
\label{sec:theoSol}
The chameleon mechanism naturally divides {the phenomenology} into two regimes.
Deep inside a massive object, where the density is high, the~field rapidly
relaxes to the minimum of $V_{\rm eff}$ and spatial gradients become negligible.
In this screened region, the~scalar field approximately satisfies
\begin{equation}
\label{eq:campoint}
\phi_{\rm int}
\simeq
\left(
\frac{Q\,\rho_{\rm tot}}
{n\,\lambda^{4+n} M_{\rm Pl}}
\right)^{-1/(n+1)},
\end{equation}
with $\rho_{\rm tot} = \sum_j \rho_j$.
As a consequence, the~fifth force is strongly suppressed and gravity effectively
reduces to GR. In~the following, as~usually done~\cite{Terukina12,Pizzuti2024b}, we will assume $\phi_{\rm int} \sim 0$.

At larger radii, where the matter density decreases and the field has not yet
settled to the minimum of the effective potential, the~contribution of $V'(\phi)$
becomes subdominant. In~this unscreened regime, the~scalar field obeys
\begin{equation}
\label{eq:luna}
\nabla^2 \phi_{\rm out}
\simeq
\frac{Q}{M_{\rm Pl}}\, \rho_{\rm tot}.
\end{equation}
Assuming spherical symmetry, which is a good approximation for relaxed galaxy
clusters~\cite{Lagan2019,Biviano:2023oyf}, this equation reduces to
\begin{equation}
\label{eq:spherical}
\frac{1}{r^2}
\frac{\mathrm{d}}{\mathrm{d}r}
\left(
r^2 \frac{\mathrm{d}\phi_{\rm out}}{\mathrm{d}r}
\right)
=
\frac{Q}{M_{\rm Pl}}\, \rho_{\rm tot}(r).
\end{equation}

A single integration shows explicitly that the scalar-field gradient, and~hence
the fifth force, is sourced by the enclosed mass profile
\begin{equation}
\label{eq:firstintegration}
r^2 \frac{\mathrm{d}\phi_{\rm out}}{\mathrm{d}r}
=
\frac{Q}{M_{\rm Pl}}
\int_0^r r'^2 \rho_{\rm tot}(r)\,\mathrm{d}r'
+ C_\text{s},
\end{equation}
where $C_\text{s}$ is an integration constant. 
The full scalar-field profile is obtained by matching the interior and exterior
solutions at a characteristic screening radius $r=S$, defined as the scale at
which the transition between the screened and unscreened regimes occurs (e.g.,~\cite{Terukina:2013eqa,Wilcox:2016guw}).
Imposing a continuity of $\phi$ and its derivative at $r=S$ uniquely fixes the
integration constant in Equation~\eqref{eq:firstintegration}.

Within $r<S$, the~field is strongly suppressed,
$\phi \ll \phi_\infty$, and~the fifth force is negligible, whereas at
$r>S$ the scalar field mediates a long-range modification of gravity. This
screening radius therefore encapsulates the environmental dependence of
chameleon and $f(R)$ gravity and plays a central role in the phenomenology of
galaxy~clusters.

Indeed, as~already pointed out in~\cite{Pizzuti2024b,Pizzuti2024c}, Equation~\eqref{eq:firstintegration} depends on the profile assumed to model the matter density distribution, impacting the shape of the chameleon field profile and  changing the efficiency of the screening mechanism. The~analysis carried out by~\cite{Pizzuti2024b} on simulated cluster-sized halos further demonstrates that the effect of mass modeling does not introduce relevant biases or spurious detections in 
 the constrained parameter space, when other systematics are under control. However, this may be not the case for real clusters, where observational systematics and assumptions in the mass reconstruction are expected to affect the {analysis} \cite{Pizzuti2026a}.

In this work, we follow the approach of~\cite{Pizzuti2026a} and we consider three mass profile ans\"atze, which have found to provide adequate fit to the total matter distributions in observed clusters~\cite{Biviano2013,Umetsu:2015baa}, namely the Navarro--Frenk--White (NFW) model~\cite{Navarro:1995iw},
\begin{equation}
        \label{eq:NFW}
         M_\text{NFW}(r) = M_{200}\frac{\text{ln}(1+r/r_\text{s}) - \frac{r/r_\text{s}}{(1+r/r_\text{s} )}  }{\text{ln}(1+r_{200}/r_\text{s}) - \frac{r_{200}/r_\text{s}}{(1+r_{200}/r_\text{s} )}  }\,,
        \end{equation}
    where $r_\text{s} \equiv r_{-2}$ is the radius at which the logarithmic derivative of the density profile {assumes a value of} $-2$;
    the Burkert model (e.g.,~Ref.~\cite{Burkert:2000di}), is defined as
\begin{equation}
         M_\text{Bur}(r) = M_{200}\frac{\text{ln}\big[1+(r/r_\text{s})^2\big] +  2\,\text{ln}(1+r/r_\text{s}) - 2\,\text{arctan}(r/r_\text{s}) }{\text{ln}\big[1+(r_{200}/r_\text{s})^2\big] +  2\,\text{ln}(1+r_{200}/r_\text{s}) - 2\,\text{arctan}(r_{200}/r_\text{s})}\,,
        \end{equation}
    where $r_\text{s} \simeq 2/3 \, r_{-2}$, and, finally, the~Hernquist profile (e.g.,~Ref.~\cite{Hernquist1990}),
\begin{equation}
         M_\text{Her}(r) = M_{200}\frac{(r_{200}+ r_\text{s})^2}{r_\text{200}^2}\frac{ r^2}{(r+ r_\text{s})^2}\,,
        \end{equation}
    with $r_\text{s} =2\, r_{-2}$. All profiles are characterized by two free parameters, the~scale radius and the ``virial'' radius $r_{200}$, defined as the radius of a sphere enclosing an average density 200 times the critical density of the universe at the cluster's redshift. The~corresponding virial mass is $M_{200} = M(r_{200})$.

    Each model exhibits a different behavior at small and large scales; for $r\to 0$ the Burkert profile flattens, while the NFW and Hernquist models diverges as $r^{-1}$. As~$r$ grows, the~ 
    Hernquist density decreases as $\sim$$r^{-4}$, faster than the other two ($\propto r^{-3}$).

    Consequently, this translates to a distinct shape for the external gradient in the last term in the RHS of Equation~\eqref{eq:gradient}. More specifically, for~the NFW model, one has
\begin{equation}
         \label{eq:NFWcham} \dr{\phi_\text{out}}=\frac{\mathcal{B}}{r_\text{s}\, x^2}\left[\frac{1}{x+1}+\ln (x+1)\right]+\frac{C_\text{s}}{r_\text{s}\, x^2}\ ,
        \end{equation}
    where $x = r/r_s$, $\mathcal{B}=\mathcal{Q}{\rhos \rs^2}/{\Mp}$, $C_\text{s}$ is the integration constant defined above and $\rhos(\rs,r_{200})$ is the typical central density parameter. The~conditions at the matching radius, $\xc = S/\rs$, are given by
\begin{equation}
         \label{eq:Burcham}\xc=\left[\frac{\mathcal{B}}{\phi_\infty}-1\right]\,,  \quad \Cs=-\phi_\infty-\mathcal{B}\ln\left(\frac{\mathcal{B}}{\phi_\infty}\right)\,. \
        \end{equation}
    Above, 
     the quantity $\phi_\infty$ represents the value of the field in the background and it is one of the free parameters we aim to constrain. The~exterior chameleon field gradient in the case of a Burkert profile is given by
\begin{equation}
             \dr{\phi_\text{out}}=  \frac{C_\text{s}}{r_\text{s}\, x^2}+\frac{\mathcal{B}}{r_\text{s}\, x^2}\Big[\frac{1}{4} \ln \left(x^2+1\right)  +\frac{1}{2} \ln (x+1)-\frac{1}{2} \tan ^{-1}(x)\Big]\ ,
        \label{eq:burket grad}
        \end{equation}
     and the matching with the inner solution is obtained when
        \begin{equation*}
         C_\text{s}=\frac{1}{4} \Big[-2 \mathcal{B} \log \left(\xc^2+1\right)+\pi \mathcal{B}-4 \phi_\infty\Big]\ .
        \end{equation*}
    To get the screening radius $S$, one has to solve  the numerical equation 
\begin{equation}
         \ln \left[\frac{\xc^2+1}{(\xc+1)^2}\right]+2 \tan ^{-1}(\xc)=\pi -4 \frac{\phi_\infty}{\mathcal{B}}\,, \label{eq:sa Burket}
        \end{equation}

    As for the Hernquist model, the~gradient reads
\begin{equation}\label{eq:dphi_bnfw}
         \frac{\text{d}\phi_\text{out}}{\text{d}r}=\frac{(x+1)^{-2} (x-3 x-1)}{2\,x^2}\mathcal{B}+\frac{\Cs}{\rs x^2}\ , 
        \end{equation}
        with the conditions
\begin{equation}\label{eq:sq bNFW} 
            \begin{split}  
             & \xc = \left(\frac{2\phi_\infty}{\mathcal{B}}\right)^{-\frac{1}{2}}-1\ , \\
             & \Cs=\frac{(1-\xc+3\,\xc)}{2(1 + \xc)^{2}}\mathcal{B}\ . \\
            \end{split}
        \end{equation}

\section{Cluster Dataset and Analysis~Set-Up}
\label{sec:data}
In this work we consider a sample of nine massive galaxy clusters from the CLASH~\cite{Postman12} and CLASH-VLT~\cite{2014Msngr.158...48R} collaborations. The~total CLASH sample includes 25 clusters for which high quality imaging data have been collected by the Hubble Space Telescope; 14 systems, accessible from the southern hemisphere, have been targeted by the spectroscopic follow up with the VIMOS instrument at the Very Large Telescope (VLT), providing a large amount---$\mathcal{O}(>10^2)$---of confirmed spectroscopic cluster members with precise velocity~determination. 

Five of the CLASH clusters have been selected for their strong lensing features, while the other 20 systems have been chosen from Chandra observations for the high X-ray temperatures ($kT_{\mathrm{X}} > 5$~keV) and morphologically regular shape. For~a large subset of the CLASH sample---including the nine clusters targeted in this work---deep ground-based multi-band imaging was obtained and analyzed for detailed weak-lensing studies as part of the CLASH program, as~presented in~\cite{Umetsu:2015baa}. The~unprecedented quality of the CLASH and CLASH-VLT dataset enabled {accurate} mass profile reconstructions using independent probes e.g.,~\,\cite{Umetsu:2015baa,Sartoris2020,Biviano:2023oyf,Bergamini_2023}. Several multi-band analyses have further addressed the dynamical and morphological properties of the CLASH clusters, e.g.,~\cite{Girardi2015,Donahue_2016,jimnenez2018,Umetsu18,Mercurio_2021,Girardi:2024pae}, which revealed complex features suggesting that some systems may be far from an equilibrium~configuration.

A series of recent works~\cite{Biviano20225_anis,Pizzuti2026a,Maraboli26a} focused on a subsample of nine CLASH clusters---Abell 383 (A383), Abell 209 (A209), RX J2129.7 $+$ 0005 (R2129), MS2137 $-$ 2353 (MS2137), RXC J2248.7 $-$ 4431 (R2248, also named Abell S1063), MACS J1931.8 $-$ 2635 (M1931), MACS J1115.9 $+$ 0129 (M1115), MACS J1206.2 $-$ 0847 (M1206), and~MACS J0329.7 $-$ 0211 (M329)---spanning the redshift range $0.18\le z\leq0.45$, to~investigate dynamical properties of member galaxies and to further constrain generic signatures of departures from~GR. 

Here we {exploit} the same nine systems to reconstruct the total gravitational potential in chameleon gravity {via} the kinematics of the member galaxies, and~to provide bounds on the background scalaron field $f_{R0}$ in the $f(R)$ sub-case. For~each cluster, we adopt an informative prior on the mass profile parameters $r_{-2}, r_{200}$ which is based on the results 
of the combined strong- and weak-lensing analyses of~\cite{Umetsu:2015baa} for all the parametric mass models discussed in Section~\ref{sec:theoSol}. This step is fundamental to break the degeneracy arising between the fifth force in chameleon screening and the parameters describing the matter distribution (see {e.g.,}~ \cite{Wilcox:2015kna}).
{The joint weak- and strong-lensing analysis of~\cite{Umetsu:2015baa} was carried out with CLUMI code~\cite{Umetsu2011,Umetsu2013}, a~likelihood-based framework formulated in terms of azimuthally averaged radial profiles of lensing observables. In~this approach, the~projected mass distribution of each cluster is represented by a piecewise-defined, radially binned convergence profile. The~analysis combines inner constraints from 16-band Hubble Space Telescope (HST) observations~\cite{Zitrin2015} with wide-field multiband imaging obtained primarily with Suprime-Cam on a Subaru Telescope~\cite{Umetsu14}. By~combining complementary lensing probes, this framework not only improves the precision of the mass reconstruction but also enables a better calibration of probe-dependent systematic effects.}

{Specifically, the~mass reconstruction was based on a joint input data vector composed of two sets of measurements: (1) central enclosed aperture-mass constraints at four equally spaced integration radii (10--40 arcsec), derived from detailed strong-lensing and weak-shear modeling of the CLASH-HST data~\cite{Zitrin2015}; and (2) weak-lensing shear and magnification-bias measurements on larger scales, evaluated in 10 logarithmically spaced radial bins. For~all profiles, the~position of the brightest cluster galaxy was adopted as the cluster centre.}

{To rigorously account for uncertainties in the reconstructed convergence profile,~\cite{Umetsu:2015baa} used a full covariance matrix that includes four distinct contributions: (1) statistical observational errors from the joint likelihood analysis; (2) residual mass-sheet uncertainty; (3) cosmic-noise covariance of an uncorrelated large-scale structure projected along the line of sight; and (4) intrinsic variations in the projected cluster lensing signal, driven primarily by halo triaxiality and correlated substructures.}

{Using the binned convergence profile and its full covariance matrix for each cluster, the~CLASH weak- and strong-lensing analysis derived the posterior distribution of the halo structural parameters for the NFW, Burkert, and~Hernquist mass models. In~the present work, we adopt the corresponding marginalized posterior distribution, denoted \(P_\mathrm{lens}(r_{-2}, r_{200})\), as~a robust, data-driven prior.}

As for the kinematic data, we consider the projected phase-space (p.p.s. hereafter) of each cluster $(R_i,v_{z,i})$, where $R_i$ is the projected position of the $i$-th member galaxy with respect to the cluster center, and~$v_{z,i}$ is the line-of-sight (los) velocity measured in the rest frame of the cluster. 
The selection of cluster members was performed by~\cite{Biviano20225_anis} using the CLUster Membership in Phase Space (\textsc{CLUMPS}) method of~\cite{Biviano21}.

We analyze the p.p.s using the \textsc{MG-MAMPOSSt} code of~\cite{Pizzuti:2020tdl},  a~version of the \textsc{MAMPOSSt} algorithm of~\cite{Mamon01} which performs a kinematic determination of the mass profile of spherical self-gravitating systems under the assumption of dynamical equilibrium. The~method is based on the Jeans equations 
 and it further assumes a Gaussian shape for the three-dimensional velocity field. \textsc{MG-MAMPOSSt} is equipped with general parameterizations of the gravitational potential describing a variety of dark matter, dark energy and modified gravity models viable at cosmological scales~\cite{Pizzuti22Vainsthein,Biviano:2023oyf}. Note that solving the Jeans equations requires knowledge of the number density distribution of the galaxies, $\nu(r)$, and~of the orbits' anisotropy profiles 
 , defined as
\begin{equation}
    \beta = 1 - \frac{\sigma^2_\theta + \sigma^2_\phi}{2\sigma^2_r}\,,
\end{equation}
where $\sigma^2_r, \sigma^2_\theta, and \sigma^2_\phi$ are the velocity dispersions along the radial, tangential, and azimuthal components respectively; in spherical symmetry, $\sigma^2_\theta = \sigma^2_\phi$. The~velocity anisotropy defines the average shapes of the orbits of galaxies in clusters, and~it is generally a function of the distance from the cluster~center. 

$\beta(r)$ cannot be observed directly due to the so-called mass-anisotropy degeneracy ({e.g.,}~\cite{BinneyMamon}); \textsc{MG-MAMPOSSt} overcomes this limitation by assuming a parametric model for the anisotropy and fitting the parameters describing $\beta(r)$ along with those of the gravitational potential.
We use the generalized Tiret model (e.g.,~\cite{Tiret01,Mamon19,Biviano20225_anis}),
\begin{equation}
         \label{eq:gT}
         \beta_{gT}(r) = \beta_0 + (\beta_\infty - \beta_0)\frac{r}{r+r_\beta}\,,
        \end{equation}
    which has been shown to provide a good description for a large variety of galaxy orbits in clusters~\cite{Mamon19}. In~Equation~\eqref{eq:gT}, $\beta_0$ and $\beta_\infty$ represent the anisotropy at $r=0$ and at very large radii, while $r_\beta$ is a scale radius and is usually set to be equal to $r_{-2}$ of the cluster mass profile~\cite{Biviano2013,Sartoris2020}.

As for the number density profile, we rely on the results of~\cite{Biviano20225_anis}, who fitted the projected distribution $N(R)$ of the member galaxies, correcting for completeness, adopting the method of~\cite{Sarazin80}. All clusters' projected number density profiles are well-described by the projected NFW model, except~for A209, for~which the King model~\cite{King1962} is favored. While both models are described by two free parameters, in~the Jeans equations, the normalization of $\nu(r)$ simplifies it, and~the only parameter left is the characteristic scale radius $r_\nu$. In~the \textsc{MG-MAMPOSSt} analysis, we consider a Gaussian prior on $r_\nu$ whose amplitude is set by the uncertainties found by the fit of~\cite{Biviano20225_anis}. A~summary of the clusters' dynamical properties and the~constraints on the mass profile parameters derived from lensing and kinematic analyses (in standard gravity), and~on the number density scale radius can be found in Table~1 of~\cite{Pizzuti2026a}.

Here we apply the same setup described in~\cite{Pizzuti2026a} and we consider galaxies within the projected region $R\in[0.05\,{\rm Mpc}, r_{200}^L]$, where $ r_{200}^L$ is the value of the virial radius estimated by the lensing analysis of Ref.~\cite{Umetsu18}. As~a consistency check, we varied the upper bounds by $\sim 10 \%$ for each cluster to~ensure {the final results are not significantly affected by this assumption}.

We implement the expression of the gravitational potential of Equation~\eqref{eq:gradient}, assuming the field profiles of Equations~(\ref{eq:NFWcham}), (\ref{eq:Burcham}) and (\ref{eq:dphi_bnfw}) for the NFW, Burkert and Hernquist models respectively. For~the general chameleon framework, the~magnitude of departures from GR is entirely specified by the coupling constant $\mathcal{Q}$ and the background field value $\phi_\infty$. However, as~discussed in Section~\ref{sec:theo}, the~shape of the fifth force also depends on the mass profile parameters. We thus perform  a Monte Carlo-Markov Chain (MCMC) exploration of the full parameter space $(r_{200},r_{s},\beta_0,\beta_\infty,r_\nu,\mathcal{Q},\phi_\infty)$. At~each trial, we sample $r_{200},r_{s}$ from the lensing distribution, while assuming flat priors on $\beta_0,\beta_\infty$. Note that, in~\textsc{MG-MAMPOSSt}, the~scaled parameters $\mathcal{A}_{0,\infty} = 1/\sqrt{1-\beta_{0,\infty}}$ are used for the anisotropy, which are always positive and $ < 1$, $=1$, and $>1$ for tangential, isotropic, and radial orbits, respectively. We consider $\mathcal{A}_{0,\infty} \in [0.4,7]$; the scale radius is instead sampled from an informative Gaussian distribution as described above.
As for the chameleon parameters, we consider the rescaled variables
(e.g.,~\cite{Wilcox:2015kna,Boumechta:2023qhd}),
$\Qtwo=Q/(1+Q)$ and $\phitwo = 1- \exp\left[\phi/(10^4\,\Mp)\right]$, which {are in the} range $[0,1]$. When focusing on the $f(R)$ class, we instead assume a flat prior in $\log_{10}f_R \in [-8,3]$.
For all the MCMC runs, we discard the first 5000 points as burn-in, and~we ensure convergence by performing a Geller–Rubin 
 test as detailed in~\cite{Pizzuti2026a}.

\section{Results}
\label{sec:results}
\begin{figure}[H]
    
    \includegraphics[width=0.99\columnwidth]{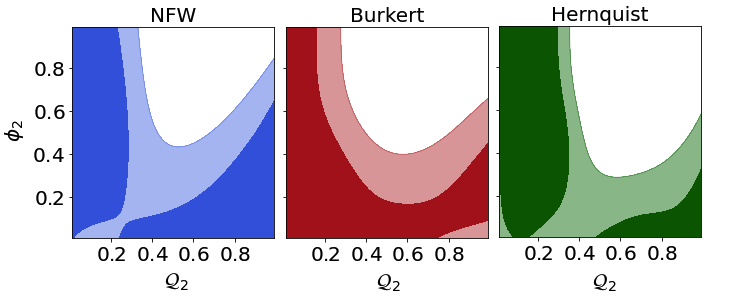}
    \caption{Two-dimensional allowed regions at $1\sigma$ (darker areas) and $2\sigma$ (lighter areas) in the $\mathcal{Q}_2,\phi_2$ plane from the \textsc{MG-MAMPOSSt} analysis of A209. (\textbf{Left}): NFW mass profile. (\textbf{Center}): Burkert profile. (\textbf{Right}): Hernquist~profile.}
    \label{fig:chamA209}
\end{figure}

We first consider a general chameleon scenario. In~Figure~\ref{fig:chamA209} we show, as~an illustrative example, the~allowed regions at $1\,\sigma$ (darker shaded areas) and $2\,\sigma$ (lighter shaded areas) in the $\mathcal{Q}_2,\phi_2$ plane obtained from the joint lensing and kinematical analysis of A209. Each panel corresponds to a different assumption for the total cluster mass~distribution.

{All the three posterior distributions in Figure~\ref{fig:chamA209} remain in agreement with the standard gravity expectation ($\mathcal{Q}_2=\phi_2=0$) at the $2\sigma$ level; nevertheless, a~marginal variation of the shape of the contours with the adopted mass profile can be observed in the regime of large deviations, $\mathcal{Q}_2 \gtrsim 0.6, \phi_2 \gtrsim 0.6$, where the excluded region slightly shrinks in the case of the NFW profile. This behavior reflects the different efficiency of the screening mechanism in clusters with distinct density slopes~\cite{Pizzuti2024b}. In~particular, the~NFW model exhibits a more efficient screening for large values of the coupling constant, generally leading to a larger allowed region in the parameter space, compared to the other two profiles.} The full set of posterior distributions for all clusters is shown in Figures~\ref{fig:allNFW}--\ref{fig:allHer} of Appendix~\ref{app:marginal}, corresponding to the NFW, Burkert, and~Hernquist mass models, respectively.

{When NFW or Hernquist profiles are assumed, no statistically significant ($\gtrsim 2\sigma$) deviation from GR is found for any cluster. A209 A383 and R2248 display mild ($\sim1\sigma$) shifts away from $\mathcal{Q}_2 = 0$, $\phi_2 = 0$ when Hernquist is assumed,  and~M329 also when the NFW model is chosen,  although~three of these systems are known to present dynamical complexity.} While the analysis of~\cite{Girardi:2024pae} revealed that M329 is not far from dynamical relaxation, the~cluster resides in a relatively rich environment; R2248 shows signs of disturbance~\cite{Mercurio_2021,Pizzuti22Vainsthein,Pizzuti2026a}, and~A209 exhibits evidence of a recent or ongoing merging event~\cite{jimnenez2018}. These factors may enhance degeneracies between dynamical mass parameters and modified-gravity degrees of~freedom.

A slightly different behavior emerges when the cluster mass distribution is modeled with the Burkert profile. In~this case, the~tension in R2248 increases to the $\sim$3$\sigma$ level, while A383, R2129 and M329 show a $\sim$1$\sigma$ preference for larger chameleon parameters. This trend is consistent with the strong+weak lensing analyses of~\cite{Umetsu:2015baa}, which slightly disfavor the Burkert model relative to cuspy profiles for our cluster sample. In~particular, for~R2248 and R2129 the lensing data {better adapt to} significantly smaller values of $r_{200}$ and $\rs$ than those inferred from kinematics in GR (see Figure~8 of~\cite{Pizzuti2026a}). 

Within the chameleon framework, part of this tension can be absorbed by the additional parameters $\mathcal{Q}$ and $\phi_\infty$. Owing to the structure of Equation~\eqref{eq:effective_potential}, the~projected phase-space distribution of member galaxies can remain nearly unchanged if one compares a deeper potential well in GR (larger $r_{200}$) with a shallower mass profile supplemented by non-zero modified-gravity parameters. Consequently, when the lensing prior favors lower halo radii, the~joint likelihood may shift probability toward {parameter space regions} associated with deviations from~GR.

To quantify whether these shifts correspond to genuine support for modified gravity, we have further computed the Bayesian evidence $\ln Z$ for each mass-profile assumption in the joint kinematic+lensing analysis within the chameleon model.  {The values of $\ln Z$ are listed in Table~\ref{tab:evidence} for all the nine clusters in the sample.}

\begin{table}[H]
\small
\caption{Log-evidence ($\ln Z$) and differences with respect to the NFW model. Positive $\Delta \ln Z$ indicates preference over~NFW.\label{tab:evidence}}
\begin{tabularx}{\textwidth}{LCCCCC}
\toprule
\textbf{Cluster} & \boldmath{$\ln Z_{\rm NFW}$} & \boldmath{$\ln Z_{\rm Hernquist}$} & \boldmath{$\ln Z_{\rm Burkert}$} & \boldmath{$\Delta \ln Z_{\rm Her}$} & \boldmath{$\Delta \ln Z_{\rm Bur}$} \\
\midrule
A209   & 8091.010 & 8088.988 & 8084.830 & $-2.022$ & $-6.180$ \\
A383   & 3771.540 & 3772.372 & 3770.583 &  $0.831$ & $-0.957$ \\
M1115  & 3303.020 & 3301.360 & 3302.977 & $-1.660$ & $-0.043$ \\
M1206  & 3297.369 & 3297.727 & 3298.346 &  $0.358$ &  $0.977$ \\
M1931  & 2035.744 & 2036.559 & 2034.657 &  $0.815$ &  $-1.08$ \\
M329   & 1797.017 & 1798.084 & 1797.753 &  $1.068$ &  $0.737 $\\
MS2137 & 1264.760 & 1264.028 & 1263.575 & $-0.732$ &  $-1.185$ \\
R2129  & 1467.501 & 1467.064 & 1465.270 & $-0.437$ & $-2.231$ \\
R2248  & 6535.780 & 6536.338 & 6532.533 &  $0.558$ & $-3.247$ \\
\bottomrule
\end{tabularx}

\end{table}

{It is worth to notice that the differences in log-evidence between NFW and Hernquist remain relatively small ($|\Delta \ln Z|\lesssim1$–$2$), suggesting that the joint lensing and kinematic data do not significantly discriminate between these cuspy parameterizations once the modified-gravity sector is included.
The Burkert profile is instead decisively disfavored in A209 ($\Delta \ln Z \simeq -6$ with respect to NFW), and~moderately disfavored in R2248 ($\Delta \ln Z \gtrsim -3$) and R2129 ($\Delta \ln Z \gtrsim -2$).} 

The evidence under {examination} thus indicates that the mild shifts toward $\mathcal{Q}_2>0$ and $\phi_2>0$ observed in some Burkert-based runs may not be robust signatures of modified gravity, but~rather manifestations of the degeneracy between mass-profile parameters and chameleon degrees of freedom under the adopted lensing priors.
Kinematic-only runs assuming flat log-priors on $r_{200}$ and $\rs${, and} shows no preference of any model, with~$\Delta \ln Z$ consistently smaller than 1. {For completeness, we also evaluated the Akaike (AIC) and Bayesian (BIC) information criteria (see e.g.,~\cite{Mamon19}), which are based on the comparison of the peak of the likelihoods and introduce penalties proportional to the number of free parameters. However, since all the mass models considered here share the same number of free parameters, these criteria reduce to a comparison based on the maximum likelihood, yielding results that are fully consistent with the Bayesian evidence.}

\begin{figure}[H]

    \includegraphics[width=0.6\columnwidth]{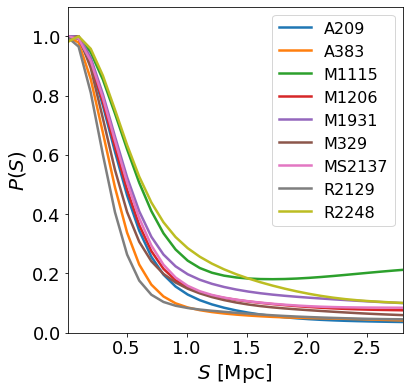}
    \caption{Posterior distributions of the screening radius $S$ obtained from the analysis of the individual clusters. An~NFW has been assumed to model the mass~distribution.}
    \label{fig:Screen}
    \end{figure}

{Figure \ref{fig:Screen} further shows the posterior distribution of the screening radius $S$ derived for each cluster assuming a NFW profile as~a function of $(r_{200},r_\text{s},\phi_2,\Qtwo)$. For~six clusters out of nine, the~distribution  decreases by 
	 90\% above $S\sim 1$ Mpc, which is in agreement with the physical scale of galaxy clusters, while R2248, M1115 and M1931 exhibit a shallower posterior distribution 
	  at large $S$ values.}

\begin{figure}[H]

    \includegraphics[width=0.99\columnwidth]{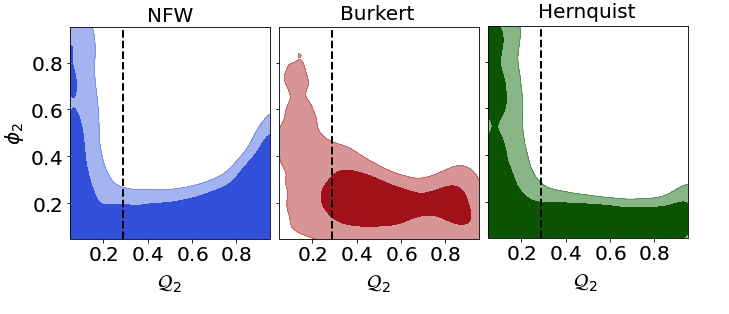}
    \caption{Two-dimensional $1\sigma$ (darker areas) and $2\sigma$ (lighter areas) allowed regions in the parameter space $\mathcal{Q}_2,\phi_2$ from the joint nine-clusters marginalized distribution. (\textbf{Left}): NFW. (\textbf{Central}): Burkert. (\textbf{Right}): Hernquist. The~vertical dashed lines refer to $\mathcal{Q} = 1/\sqrt{6}$ (i.e.,~the sub-case of chameleon $f(R)$).}
    \label{fig:Chamjoint}
    \end{figure}

In principle, the~background chameleon field evolves with cosmological time {({i.e.,}:} $\phi_\infty = \phi_\infty(z){)}$; this evolution is driven by the parameters characterizing the field potential ({e.g.,}~\cite{Brax:2004qh,Mota_2012,Burrage:2017shh}); however, given the limited redshift range (0.19--0.45) spawned by our cluster sample, we can {safely} neglect the time evolution and combine the independent distributions of $\phitwo,\Qtwo$ of all clustera together. The~joint two-dimensional constraints for $\phitwo$ and $\Qtwo$ are shown in Figure~\ref{fig:Chamjoint} for the three mass~models.
The NFW and Hernquist models show full agreement with GR, further providing strong bounds on the allowed parameter space, excluding the region $\phitwo,\Qtwo \gtrsim 0.4$. This confirms the forecasts of~\cite{Pizzuti:2020tdl,Pizzuti2024b} and it results in the most stringent constraints on the general chameleon model at Mpc scale using galaxy clusters so~far. 

On the other hand, the~Burkert model leads to a $\sim$2$\sigma$ departure from the GR expectation in the combined analysis of all nine {objects}. This apparent tension originates from the slight cluster-by-cluster shifts in the posterior distribution 
 towards $\phi_2,\mathcal{Q}_2 > 0$, which adds coherency to 
  the joint~constraint.

\subsection*{The $f(R)$ Sub-Case}
\begin{figure}[H]
        
        \includegraphics[width=0.99\columnwidth]{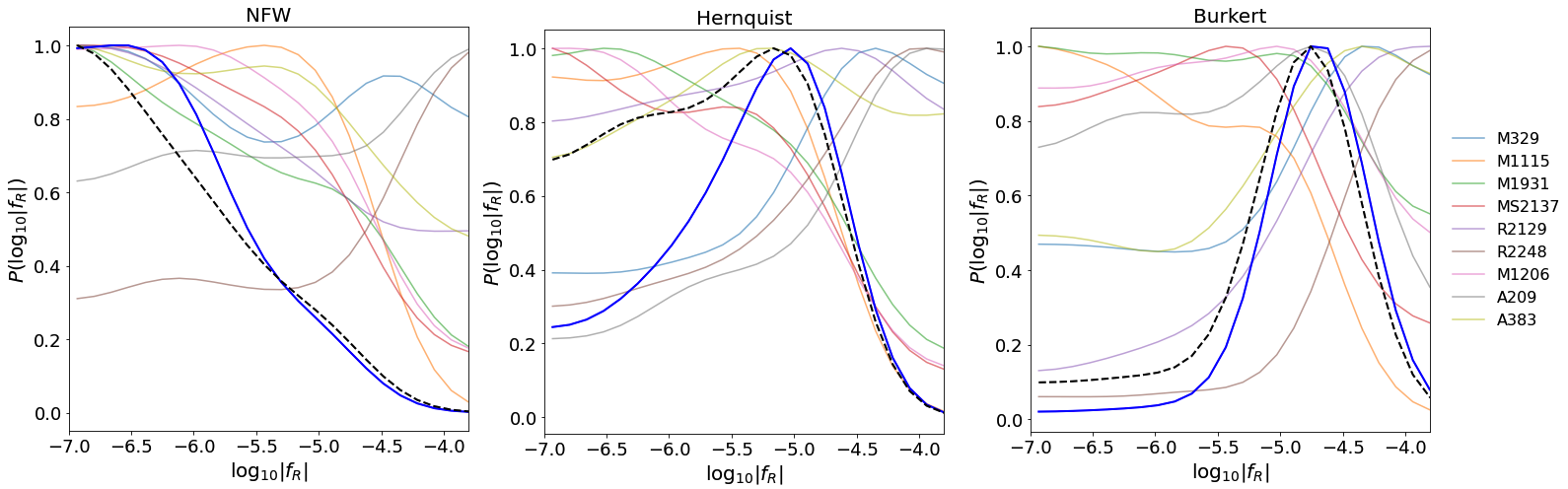}
        \caption{Marginalized distributions of the $f_R$ background field from the \textsc{MG-MAMPOSSt} kinematic + lensing analysis of the nine clusters in our sample. (\textbf{Left}): NFW. (\textbf{Central}): Hernquist. (\textbf{Right}): Burkert. The~solid blue lines correspond to the joint distributions obtained considering all the clusters, while the dashed black lines refer to the case where disturbed clusters are~excluded.}
        \label{fig:allfR}
    \end{figure}

As a second step, we consider the $f(R)$ class of chameleon models by fixing $\mathcal{Q} = 1/\sqrt{6}$ and identifying the scalaron field $f_R$ as a function of the chameleon field $\phi$ according to Equation~\eqref{eq:conversion}\endnote{To avoid excess notation, from~now on, we indicate with $f_R$ the background scalaron field.}{.} 
  We ran the \textsc{MG-MAMPOSSt}
analysis again on all the clusters in our sample; the resulting one-dimensional distributions of $\log_{10}|f_R|$ are shown for NFW, Burkert and Hernquist in the three plots of Figure~\ref{fig:allfR}. In~each panel we depict the single-cluster posteriors (solid lines with partial transparency) and the joint distribution obtained by combining the nine marginalized posteriors (solid line with full opacity).

Similarly to what found for the general chameleon framework, the~joint posteriors for NFW and Hernquist mass models are in agreement with $|f_R|\to 0$, despite a slight $\sim 1\sigma$ shift in the {latter} case.

\begin{figure}[H]
        
        \includegraphics[width=0.8\columnwidth]{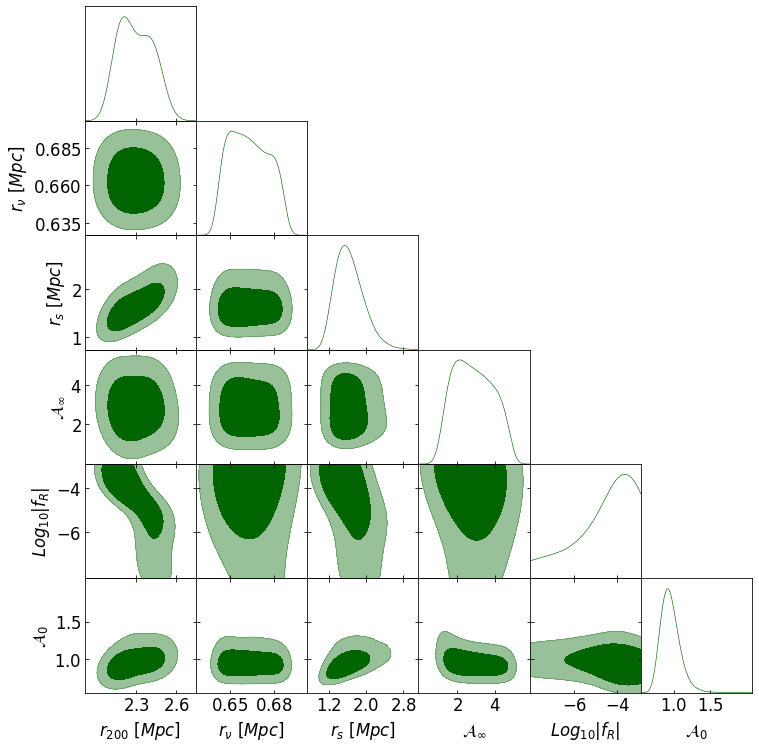}
        \caption{{Marginalized one-dimensional (darker filled regions) and two-dimensional (lighter filled regions)} 
 distributions of free parameters in the \textsc{MG-MAMPOSSt} kinematic + lensing analysis of A209 in $f(R)$ gravity. The~Hernquist model has been assumed for the total mass~profile.}
        \label{fig:A209all}
    \end{figure}

In Figure~\ref{fig:A209all}, we further show an example of the triangle plots for all the parameters involved in the analysis of A209 with the Hernquist profile. The~shapes of the contours reflect the degeneracy among the scalaron field and the mass profile parameters $r_{200},\rs$. In~particular, the~two-dimensional plane $r_{200}-\log_{10}|f_R|$ shows that the effect of the fifth force can be compensated by a shallower potential well (i.e.,~allowing for smaller $r_{200}$).

Overall, we obtain a joint constrain of $|f_R| \lesssim 2\times 10^{-5}$ and $|f_R| \lesssim 5\times 10^{-5}$ at 95\% confidence limit for NFW and Hernquist, respectively.

As expected, the~systematic tension with GR is more pronounced when adopting the Burkert profile, leading to a combined constraint of
\begin{equation}
\log_{10}|f_R| = -4.7 \pm 1.2 ,
\end{equation}
which is in clear tension with current astrophysical and cosmological bounds (e.g.,~\cite{Tamosiunas_2022,Bai25}).

To support the idea that this apparent deviation is driven by residual systematics rather than by genuine modified-gravity signatures, we follow the procedure outlined in~\cite{Pizzuti2026a} and restrict the joint analysis to dynamically relaxed systems. Specifically, we exclude clusters showing clear evidence of dynamical disturbance, as~quantified by two complementary indicators: the fraction of galaxies in substructures, $f_g$, and~the Anderson--Darling statistic $A^2$ computed from the line-of-sight velocity distribution of member~galaxies.

The $A^2$ parameter measures departures from Gaussianity in the velocity distribution and has been shown to correlate strongly with the likelihood of spurious modified-gravity signals induced by non-equilibrium dynamics~\cite{Pizzuti2020syst}. The~fraction of galaxies in substructures, $f_g$, was estimated by~\cite{Biviano20225_anis} using the DS+ algorithm developed by~\cite{Benavides23}, which exploits the full p.p.s. information to identify substructures. The~corresponding values of $A^2$ and $f_g$ are reported in the last two columns of Table~1 of~\cite{Pizzuti2026a}, and~we will not repeat them~here. 

{Note that selecting clusters with low values of $A^2$ and $f_g$ does not only isolate dynamically relaxed systems, but~also indirectly favors systems with a more regular and symmetric mass distribution. Indeed, relaxed clusters are expected to be closer to spherical symmetry, while strongly disturbed systems are more likely to exhibit significant triaxiality and projection effects.}

{This consideration is further supported by the properties of the CLASH sample itself, which has been shown from dedicated cosmological hydrodynamical simulations to not exhibit a significant orientation bias~\cite{Meneghetti_2014}, making the use of spherical models a reasonable approximation at the level of precision considered here. Therefore, this selection partially mitigates potential systematics associated with departures from spherical symmetry. It is also important to point out that triaxiality is already considered as a source of statistical uncertainties in the strong+weak lensing covariance matrices, calibrated on numerical simulations~\cite{Umetsu14,Umetsu:2015baa}}

We remove from the joint fit the systems with $f_g > 0.3$ and $A^2 >0.7$, namely A209, M1115, R2248 and MS2137; the resulting distributions are shown as the black dashed lines in Figure~\ref{fig:allfR}.
While the upper limits are almost unchanged, a~rise in the marginalized distribution at $|f_R|\to 0$ is found in all the cases, most notably for the Hernquist mass model.
This exercise stresses once again the importance of systematic calibration and accurate mass modeling when using cluster mass profiles to test the nature of gravity.
 The fact that our MG constraints remain stable when restricting the analysis to this sub-sample suggests that residual triaxiality is not the dominant driver of the observed trends, and~that the main systematic effects are instead related to dynamical~non-equilibrium.

\section{Conclusions}
\label{sec:conclusions}

In this work we have performed a joint kinematic--lensing test of chameleon screening gravity, and~of its $f(R)$ sub-class, using nine massive galaxy clusters drawn from the CLASH and CLASH-VLT datasets. We modeled the dynamics of member galaxies with the \textsc{MG-MAMPOSSt} code under the assumptions of spherical symmetry and dynamical equilibrium, and~we adopted informative priors on the halo structural parameters $(r_{200},\rs)$ from strong+weak lensing reconstructions~\cite{Umetsu:2015baa} (as implemented in~\cite{Pizzuti2026a}). 

For the general chameleon scenario, we explored the parameter space
$(\mathcal{Q},\phi_\infty)$ adopting three alternative ans\"atze for the total mass distribution (NFW, Burkert, and~Hernquist). When cuspy mass profiles are assumed (NFW or Hernquist), the~combined nine-cluster constraints are consistent with GR and exclude large regions of the $(\mathcal{Q}_2,\phi_2)$ plane, roughly ruling out $\mathcal{Q}_2,\phi_2 \gtrsim 0.4$ at high credibility. These results are broadly consistent with expectations from previous cluster-scale forecasts~\cite{Pizzuti:2020tdl,Pizzuti2024b} and analyses of real samples~\cite{Terukina:2013eqa,Wilcox:2015kna,Boumechta:2023qhd} and show that joint lensing--dynamics {investigations} can deliver stringent constraints on screened modifications of gravity at Mpc~scales.

A different behavior emerges when adopting a Burkert profile for the total mass distribution. In~this case, the~cluster-by-cluster posteriors exhibit mild coherent shifts toward $\mathcal{Q}_2>0$ and $\phi_2>0$, which add up in the combined constraint and yield an apparent $\sim$2$\sigma$ departure from the GR expectation. The~evidence analysis indicates that these shifts do not constitute robust support for modified gravity, but~are more naturally interpreted as a manifestation of the increased degeneracy between the modified-gravity sector and the mass-profile parameters when the assumed inner slope is in tension with the lensing constraints for part of the~sample.

We then focused on the $f(R)$ sub-case by fixing $\mathcal{Q}=1/\sqrt{6}$ and sampling the logarithm of the scalaron field $\log_{10}|f_R|$ with a flat, uninformative prior. For~NFW and Hernquist, we find joint constraints consistent with $|f_R|\to 0$, and~we obtain upper limits at the level of $|f_R|\lesssim 2-5\times 10^{-5}$ (95\% C.L.), in~agreement with current cosmological constraints. Conversely, assuming a Burkert profile leads to a combined posterior peak 
 away from zero,
$\log_{10}|f_R| = -4.7\pm 1.2$, which is in tension with astrophysical and cosmological bounds ({e.g.,}\ \cite{Tamosiunas_2022,Bai25}).

We have explicitly investigated the role of systematics in the dynamical state of the clusters by repeating the analysis after excluding clusters that show clear indications of non-equilibrium, based on the substructure fraction $f_g$ and the Anderson--Darling statistic $A^2$. This {procedure} {led to} a shift of the marginalized posteriors toward $|f_R|\to 0$ for all mass models, supporting the interpretation that part of the apparent Burkert-driven tension is induced by residual dynamical complexity of the systems. At~the same time, this exercise should be regarded as a targeted (and intentionally conservative) mitigation strategy rather than as an exhaustive treatment of systematics: additional effects such as projection effects and baryonic feedback are known to influence lensing mass determinations~\cite{Grandis_2021,Giocoli24}, see also the review of~\cite{Umetsu2020}; furthermore, residual l.o.s. interlopers contamination and potential modeling limitations of simple two-parameter mass profiles may still affect the inference at a non-negligible level, demanding a more accurate, multi-component analysis ({e.g.,}~\cite{Biviano:2023oyf,Pizzuti2024c}). {Finally, it has been shown that triaxiality---although not being the dominant sources of systematics---may boost the mass profile when observing clusters along the line of sight~\cite{Pizzuti2020syst}. As~concerns lensing analyses, if~the sample of clusters suffers an orientation bias toward alignment with the line of sight, then assuming either spherical or elliptical profiles would still lead to a net bias in the mass estimates~\cite{Umetsu18}. For~this reason, a~dedicated upcoming work is addressing the extension of our framework to triaxial configurations of the mass profiles.}

 Note that strengthening cluster-scale tests of gravity requires both a broadening of the statistical sample (i.e.,~more clusters) and a meticulous control of systematics through a more comprehensive multi-probe approach. This will become increasingly feasible thanks to the combination of new-generation wide-field {{photometric}
} surveys enabling precise weak-lensing mass measurements---such as \emph{{Euclid}} \cite{EuclidSkyOverview}, the~Vera Rubin Observatory~\cite{Ivezic2019}, and~the Roman Space Telescope~\cite{sanderson2024recommendationsearlydefinitionscience}---and dedicated {{spectroscopic}} follow-up programs aimed at measuring member-galaxy velocities for large cluster samples (e.g.,~DESI~\cite{DESI21}, 4MOST~\cite{4MOST2019}, or~MOONS/VLT~\cite{Cirasuolo2020}). In~addition, several cluster-dedicated programs are already providing high-quality, multi-wavelength mass reconstructions that are particularly well suited for gravity tests at halo scales. Examples include X-COP~\cite{Eckert2017}, CHEX-MATE~\cite{chexmate21}, and~HeCS/HeCS-SZ~\cite{Rines_2013,Rines_2016}, which combine X-ray, Sunyaev--Zel'dovich, and~optical spectroscopic data to deliver accurate determinations of thermodynamic and dynamical profiles from the cluster core to the outskirts. Extending joint lensing--kinematics analyses to $\mathcal{O}(10^2)$--$\mathcal{O}(10^3)$ systems will not only reduce statistical uncertainties, but~will also enable stringent consistency tests across dynamical state, mass, redshift, and~environment, thereby isolating (and potentially modeling) the dominant sources of residual~systematics.

Finally, while chameleon screening and simple $f(R)$ parameterizations are nowadays very well constrained, they remain an important and well-defined theoretical laboratory of modified gravity at cluster scales, offering a valuable baseline for more complex scenarios (e.g.,~extended scalar--tensor models~\cite{2016JCAP...04..044C}, environment-dependent couplings, or~beyond-Horndeski~\cite{Kobayashi_2019, Gleyzes_2015, Zumalac_rregui_2014}/degenerate higher-order theories~\cite{2017JCAP...05..033L, Langlois2019}) in which the phenomenology at cluster scales may be richer. In~this sense, the~present analysis represents a methodological stepping stone toward more general tests of gravity with upcoming cluster~datasets.

\vspace{6pt}

\authorcontributions{Conceptualization, L.P.; methodology, L.P.;
software, L.P.; validation, L.P., K.U.
and A.B.; formal analysis, L.P.; investigation,
L.P. and F.R.; resources, K.U. and
A.B.; data curation, K.U.; writing---original draft,
L.P.; writing---review and editing, L.P., F.R., K.U. and A.B.; supervision, L.P.;
Project administration, L.P. {All authors have read and agreed to the published version of the manuscript.} 
}

\funding{{L.P. acknowledges the support by the Italian Ministry for Research and University (MUR) under Grant `Progetto Dipartimenti di Eccellenza 2023-2027' (BiCoQ). K.U. acknowledges support from the National Science and Technology Council, Taiwan (grant NSTC 112-2112-M-001-027-MY3) and the Academia Sinica Investigator award (grant AS-IA-112-M04).}}

\dataavailability{Part of the data are already publicly available on the CLASH-VLT database. Kinematic data for some clusters analyzed here have not been published yet and they are available upon reasonable request from the CLASH/CLASH-VLT team.}

\acknowledgments{{The} 
 authors acknowledge the CLASH-VLT team and its PI P. Rosati for the dataset provided to carry out the present work. We also thank the anonymous referees for the valuable comments and suggestions which helped improving our work.

}

\conflictsofinterest{The authors declare no conflicts of interest.}

\appendixtitles{yes} 
\appendixstart
\appendix
\section[\appendixname~\thesection]{Marginal Distributions for the General Chameleon~Case}\label{app:marginal}

\begin{figure}[H]

    \includegraphics[width=0.99\columnwidth]{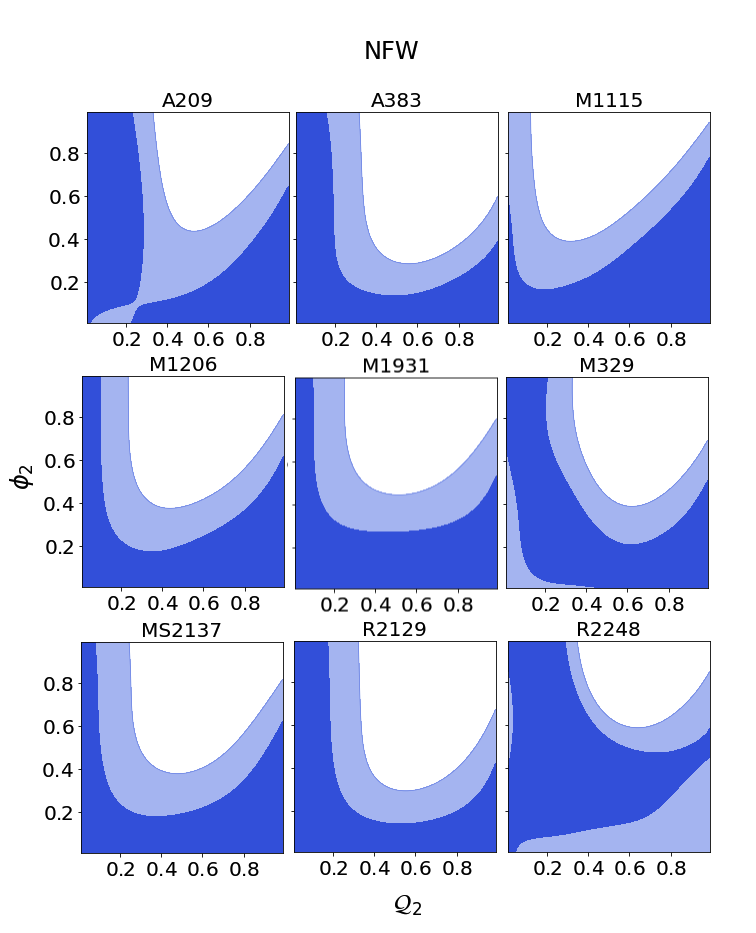}
\caption{Two-dimensional $1\sigma$ (darker areas) and $2\sigma$ (lighter areas) allowed regions in the parameter space $\mathcal{Q}_2,\phi_2$ from the \textsc{MG-MAMPOSSt} kinematic + lensing analysis of the nine clusters in our sample. The~NFW model has been assumed for the total matter~distribution.}
\label{fig:allNFW}
\end{figure}

\begin{figure}[H]

    \includegraphics[width=0.99\columnwidth]{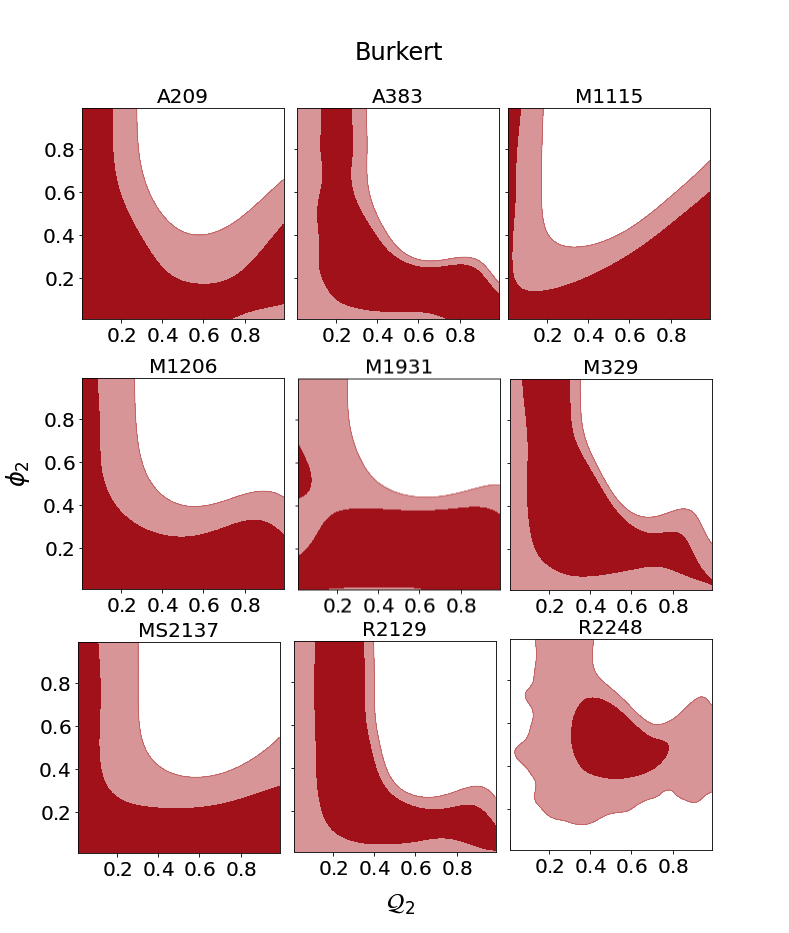}
    \caption{Same as Figure~\ref{fig:allNFW} but for the Burkert mass~model.}
    \label{fig:allBur}
    \end{figure}

\begin{figure}[H]

    \includegraphics[width=0.99\columnwidth]{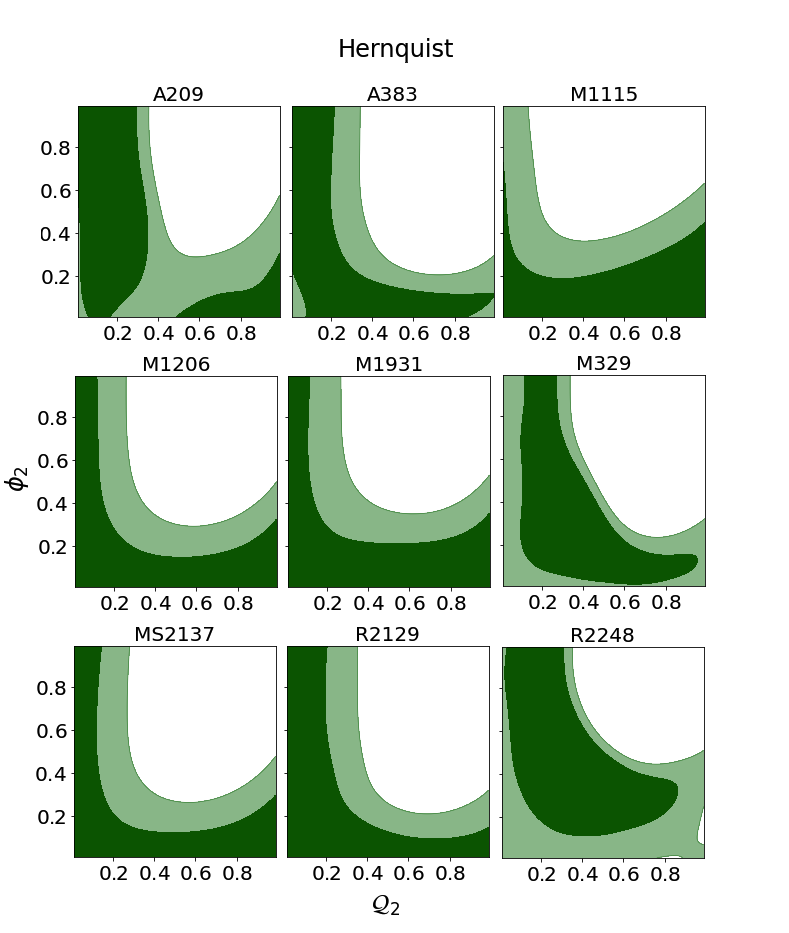}
    \caption{Same as Figure~\ref{fig:allNFW} but for the Hernquist mass~model.}
    \label{fig:allHer}
    \end{figure}

     Figures~\ref{fig:allNFW}--\ref{fig:allHer} show the two-dimensional marginalized distributions at $1\sigma$ and $2\sigma$ (darker and lighter shaded areas) for all the clusters in our sample. Each figure presents the results for a different mass model used to describe the cluster matter profile in the \textsc{MG-MAMPOSSt} fit.

\unskip

\unskip

\begin{adjustwidth}{-\extralength}{0cm}
\printendnotes[custom]

\reftitle{References}


\PublishersNote{}

\end{adjustwidth}

\end{document}